\documentstyle[11pt,newpasp,twoside,epsf]{article}
\def\edcomment#1{\iffalse\marginpar{\raggedright\sl#1\/}\else\relax\fi}
\reversemarginpar

\begin{document}
\title{Neutral hydrogen absorption at the center of NGC~2146}
 \author{A. Tarchi$^{1,2}$, A. Greve$^{3}$, A.B. Peck$^{4}$, N. Neininger$^{5}$, K.A. Wills$^{6}$, A. Pedlar$^{7}$, and U. Klein$^{5}$}
\affil{$^{1}$Instituto di Radioastronomia, CNR, Bologna, Italy}
\affil{$^{2}$Osservatorio Astronomico di Cagliari, Capoterra (CA), Italy}
\affil{$^{3}$IRAM, 300, St. Martin d'H{\`e}res, France}
\affil{$^{4}$Harvard-Smithsonian CfA, SMA Project, Hilo, HI, USA} 
\affil{$^{5}$Radioastronomisches Institut der Universit{\"a}t Bonn, Germany}
\affil{$^{6}$Department of Physics and Astronomy, Sheffield, UK}
\affil{$^{7}$Jodrell Bank Observatory, Macclesfield, Cheshire, UK}

\begin{abstract} 
We present 1.4 GHz {\mbox{H\,{\sc i}}} absorption line observations towards
the starburst in NGC~2146, made with the VLA and MERLIN. The {\mbox{H\,{\sc i}}} absorption
has a regular spatial and regular velocity distribution, and does
not reveal any anomaly as a sign of an encounter with another galaxy or of
a far-evolved merger.
\end{abstract}

\section{Scientific background}
It is well established that NGC~2146 is undergoing a strong starburst, even 
stronger than that in M\,82 (e.g.\ Kronberg \& Biermann 1985; Tarchi et al.\ 2000). However, the origin of this starburst is 
still unclear. Different hypotheses have been proposed, although without compelling evidence:\\
i) a collision with another galaxy (Young et al.\ 1988)\\
ii) a fairly gentle far-evolved merger event (Hutchings et al.\ 1990)\\
iii) a tidal interaction with a Low Surface Brightness companion (Taramopoulos et al.\ 2001)

In order to study the kinematics and density of gas in the central region of 
NGC~2146 in the light of the proposed merger/encounter hypotheses, we have 
mapped the {\mbox{H\,{\sc i}}} absorption towards the nuclear radio continuum emission using the VLA (1\farcs8 res.\ = 130\,pc) and MERLIN (0\farcs2 res.\ = 15\,pc). At the distance of NGC~2146 (14.5\,Mpc), 1$''$ is equivalent to 70\,pc.

\section{Results and Conclusions}
{\bf {VLA}} The VLA A--array observation presented here reveals {\mbox{H\,{\sc i}}} absorption in front of the radio continuum emitted in the central $\sim$ 2 kpc of NGC~2146 (Fig.~1{\bf {a}}), allowing the construction of a map of its spatial distribution and its rotation.\\
The optical depth lies between a minimum value of 0.3 and a maximum of 0.9, which for a typical linewidth of 100 {km s$^{-1}$} corresponds to {\mbox{H\,{\sc i}}} column densities between 6 and 18 $\rm \times 10^{21}\:atoms\:cm^{-2}$, respectively.\\
The {\mbox{H\,{\sc i}}} absorption velocity field is very smooth, similar to the rotation of the molecular and ionized gas as shown in Fig.~1{\bf {b}}, which supports the structure of a rotating disk of {\mbox{H\,{\sc i}}} gas. The {\mbox{H\,{\sc i}}} absorption has a uniform spatial and velocity distribution, and does not reveal any anomalous concentration or velocities which might indicate an encounter with another galaxy or a far--evolved merger.\\

\begin{figure}
\plotone{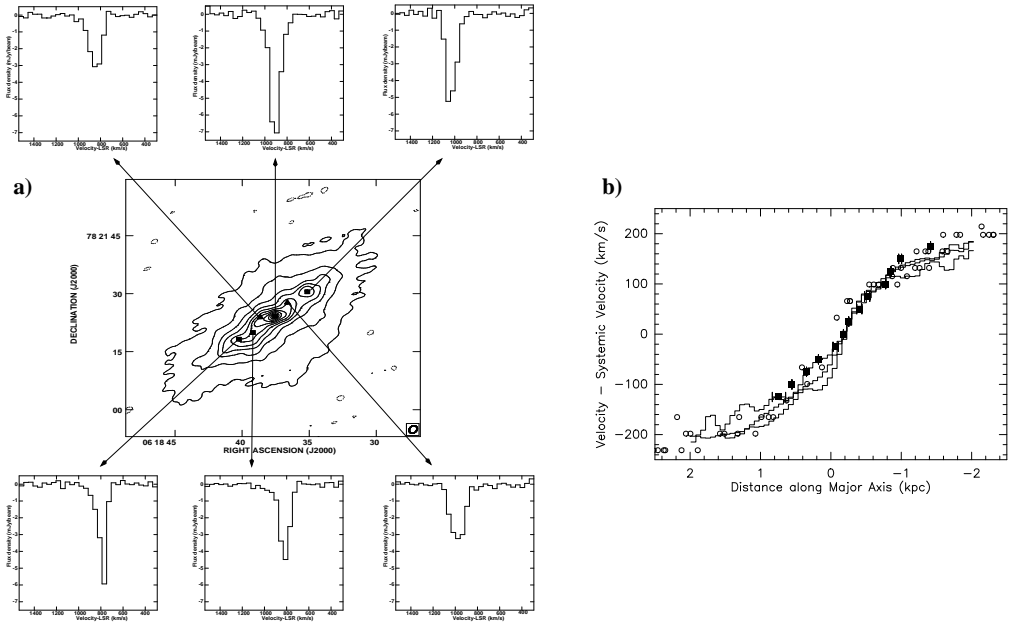}
\caption{{\bf {a)}} Naturally weighted 1.4\,GHz VLA image of NGC~2146 (big panel) and {\mbox{H\,{\sc i}}} absorption--line velocity profiles (small panels) obtained at six positions. {\bf {b)}} Position--velocity diagrams taken along the major axis of NGC~2146. Histogram lines: $^{12}$CO(1--0), (2--1) and $^{13}$CO(1--0); open circles: H$\alpha$ from Benvenuti et al. (1975); filled squares: {\mbox{H\,{\sc i}}} absorption.}
\end{figure}

\noindent {\bf {MERLIN:}} In the 1.4\,GHz naturally weighted continuum MERLIN map we find 10 of the 18 compact sources detected at 5\,GHz as reported by Tarchi et al.\ (2000). Of these 10 sources only the central one shows a clear {\mbox{H\,{\sc i}}} absorption line. The absorption properties derived from this line are in good agreement between the VLA and MERLIN data. The other 1.4\,GHz MERLIN detected sources are too weak at the MERLIN resolution, and do not show any absorption above the noise level.

\acknowledgments
We are grateful to the VLA and MERLIN staff for technical support. We would like to thank Daniela Vergani and Filippo Fraternali for insightful discussions, and Tom Muxlow and Simon Garrington for useful comments.

\end{document}